\pageno=0 
\baselineskip=15pt plus 2pt
\magnification=\magstep1 
\hsize=5.9truein
\vsize=8.7truein
\hfuzz=10pt 
\def \e{{\rm e}} 
\def \m{\textstyle {1\over 2}} 
\null  
\centerline{ \bf RAPIDITY GAPS AND PRODUCTION OF MINIJETS} 
\centerline{ \bf IN HIGH-ENERGY HADRONIC COLLISIONS} 
\vskip .25in 
\centerline{G. Calucci, R. Ragazzon and D. Treleani} 
\vskip .05in 
\centerline{\it Dipartimento di Fisica Teorica dell'Universit\`a  
and INFN Sezione di Trieste} 
\centerline{\it Trieste, I 34014 Italy} 
\vfill 
\centerline{ABSTRACT} 
\vskip .25in 
\midinsert 
\narrower\narrower  
\noindent   
High energy hadronic interactions can produce a final 
state characterized by minijets separated by a large gap in  
the rapidity distribution of the produced secondary particles.  
We discuss the process by keeping into account the possibility of having
multiple parton collisions in the hadronic interaction. At Tevatron
energy the correction to the single scattering term induced by the presence of
multiparton interactions is large for transverse momenta
smaller than $6GeV$.
\endinsert  
\vfill\eject 
\par	{\bf 1. Introduction} 
\vskip.25in 
\par 
In the kinematical regime of 
semihard hadronic collisions
the momentum transfer $q_t$ is large enough to apply perturbation 
theory but it is kept fixed with the c.m. energy, in such a way  
that the Regge limit $t/s\to0$ is reached from the perturbative 
side. The main feature, which becomes more and 
more apparent when approaching the Regge limit, is the increasing 
complexity of the process. In fact at large $q_t$  
several major aspects of the interaction are described 
by a single partonic collision represented with a Feynman diagram 
at the lowest order in the coupling constant.  
When moving towards the semihard regime the  
Regge limit is approached both by the hadronic 
and by the typical partonic collision.  
As a consequence the partonic process is not well represented  
any more by means of a tree level Feynman diagram.  
Partonic interaction in the Regge limit have been investigated 
extensively[1], [2]. An approach which has received 
a lot of attention is the BFKL Pomeron[2]: the partonic 
reaction is described by the exchange of a gluon 
ladder, with vacuum quantum numbers in the  
$t$-channel, which is constructed by neglecting   
the transverse momentum components when compared to the longitudinal 
ones. As a result of the simplified kinematics the steps of the ladder are  
ordered in rapidity and are build up with 
two basic elements: 
\item{a-} 
the gauge independent non-local vertices, which sum up the 
dominant term, in the 
$t/s\to0$ limit, of the diagrams with 
gluon emission from all near-by lines, and  
\item{b-}  
the Reggeization of the $t$-channel gluons, which is introduced
in order to keep into account 
the leading virtual corrections and which 
allows a solution to the infrared problem. 
\par\noindent 
The iteration of the ladder in the $t$-channel, is expressed 
as an integral equation (Lipatov's equation) which can be solved  
explicitly. By considering the parton process represented with a cut  
BFKL Pomeron one may write 
explicitly the expression for the inclusive cross section to produce 
minijets, which are to be identified with final state 
partons with transverse momentum  
larger than the lower threshold of observability $q_t^{min}$. 
One may also consider the elastic parton interaction with BFKL Pomeron 
exchange.  
The elastic parton collision gives rise to a distinct 
signature in the final state generated by the parton process: two minijets  
are produced with no further particles in the rapidity interval between 
them. 
On the contrary in the typical inelastic parton collision, represented with 
a cut BFKL Pomeron, the gap is filled uniformly on the average by gluons. The 
dependence of the parton cross section on the rapidity gap is 
predicted by the perturbative calculation and it is related  
by unitarity to the dependence 
of the inelastic process on the width of the rapidity interval.  
The identification of these 
features in a semihard hadronic process would signal the 
underlying parton dynamics in a distinctive way[3]. 
A difficulty which has been pointed out is that  
the hadronic event which contains the semihard partonic interaction 
is going to fill the gap in most of the cases. The perturbative 
cross section with rapidity gap has therefore been multiplied by a survival  
probability factor[4]. The survival probability has been recently estimated 
by taking explicitly into account the underlying soft hadronic event  
and it turns out to be roughly constant  
as a function of the rapidity gap[5]. The underlying event would 
therefore provide only a rescaling factor to the cross section 
calculated perturbatively and the behaviour of the actual experimental 
cross section as a function of the rapidity gap 
would still be linked directly to the BFKL dynamics.  
\par 
While testing the validity of the BFKL approach to semihard  
parton dynamics is presently 
one of the main topics in perturbative QCD, the delicate point is to 
keep properly into account the structure of the 
whole hadronic interaction, whose effect 
may mask the BFKL dynamics[6]. In fact the BFKL 
regime requires $q_t^{min}$ small with respect to all longitudinal 
momenta and smaller values of 
$q_t^{min}$ correspond to larger values for the partonic cross 
section. In the BFKL regime unitarity corrections are therefore important. 
Indeed the closer is parton dynamics  
to the BFKL limit the stronger is the effect of unitarization[6]. 
\par 
In the present paper we discuss the effect of the unitarization of the 
semihard hadronic interaction on the cross section for minijet production 
with rapidity gaps in the distribution of final state secondaries.  
The unitarization of the hadronic semihard interaction induces multiple  
semihard partonic collisions in the inelastic event 
and the cross section to produce minijets with associated 
rapidity gap is therefore modified. 
To keep into account multiple BFKL Pomeron exchanges we assume  
the validity of the AGK cutting rules for semihard interactions. 
As a consequence 
the whole semihard hadronic process is represented as a 
probabilistic superposition of multiple BFKL Pomeron exchanges. 
\par\noindent
The paper is organized as follows: In the next paragraph 
the single parton interaction is discussed. The argument of the
following section is multiple parton collisions; we describe the
general formulation to the problem and we derive the 
expression for the cross section with rapidity gap in the simplest case 
of multiple parton interactions. 
A numerical example and some concluding remarks are the 
argument of the last section while a few possible generalizations are discussed in the appendix.
\vskip.25in 
\par	{\bf 2. Single scattering term} 
\vskip.25in 
\par 
By considering the cut BFKL Pomeron one may write 
the expression for 
the cross section where two gluons interact producing many gluons  
and two of them, the ones with largest rapidity (in absolute value)
in the overall c.m. frame, 
are observed. If $y$ is the separation in rapidity and $k_a$, $k_b$  
are the transverse momenta 
of the observed gluons, the inclusive cross section is expressed as 
 
$${d\hat{\sigma}_L\over d^2k_ad^2k_b}=\biggl[{C_A\alpha_s\over 
k_a^2}\biggr] 
     f(k_a,k_b,y)\biggl[{C_A\alpha_s\over k_b^2}\biggr]\eqno(1)$$ 
 
\par\noindent 
where $C_A=N_c$ is the number of colors, $\alpha_s$ is the strong 
coupling 
constant and $f(k_a,k_b,y)$ is the inverse Laplace transform of the  
solution to Lipatov's 
equation. Actually: 
 
$$f(k_a,k_b,y)={1\over(2\pi)^2k_ak_b}\sum_{{\rm n}=-\infty}^{+\infty} 
   e^{i{\rm n}\phi}\int_{-\infty}^{+\infty}d\nu e^{\omega(\nu,{\rm n})y} 
   e^{i\nu{\rm ln}(k_a^2/k_b^2)}\eqno(2)$$ 
 
\par\noindent 
where $\phi$ is the azimuth angle between the observed gluons, 
 
$$\omega(\nu,{\rm n})=-2{\alpha_sN_c\over\pi}\Re 
    \biggl[\psi\Bigl({\mid{\rm n}\mid+1\over2}+i\nu\Bigr)-
\psi(1)\biggr]\eqno(3)$$ 
 
\noindent 
and 
 
$$\psi(z)={d{\rm ln}\Gamma(z)\over dz}\eqno(4)$$ 
 
\noindent 
is the Digamma function. The inclusive cross section for production 
of two minijets, as a result of cutting the exchange 
in the forward direction 
of a BFKL Pomeron, is obtained by folding Eq.(1) with the structure 
functions of the interacting hadrons $A$ and $B$: 
 
$${d\sigma_L\over dx_Adx_Bd^2k_ad^2k_b}= 
  f_{eff}(x_A,k_a^2)f_{eff}(x_B,k_b^2) 
  {d\hat{\sigma}_L\over d^2k_ad^2k_b}\eqno(5)$$ 
 
\noindent 
where $f_{eff}$ is the effective structure function 
 
$$f_{eff}(x)=G(x)+{4\over 9}\sum_f\Bigl[Q_f(x)+{\bar Q}_f(x)\Bigr] 
  \eqno(6)$$ 
 
\noindent 
namely the gluon structure function plus $4/9$ of the quark  
and anti-quark structure functions with flavor $f$.  
In the BFKL dynamics one can relate the rapidity $y$ of the minijet 
which carries most of the momentum of the 
initial state parton (primary minijet) with the fractional momentum  
variable $x$ of the incoming parton. The relation   
is $x=k_t{\rm e}^{y}/\sqrt{s}$ for forward final 
state partons with transverse momentum $k_t$ and rapidity $y$ and 
$x=k_t{\rm e}^{-y}/\sqrt{s}$ for backward partons. 
Eq.(1) can be integrated on the transverse momenta down to the 
lower cut off $q_t^{min}$: 
 
 $$\hat{\sigma}_L(y)=\int_{q_t^{min}}{d\hat{\sigma}_L\over 
d^2k_ad^2k_b} 
  d^2k_ad^2k_b= 
  \Bigl({\alpha_sC_A\over\pi}\Bigr)^2 
  {\pi^3\over2(q_t^{min})^2}F_L(y)\eqno(7)$$ 
 
\par\noindent 
where 
 
$$F_L(y)=\int{d\nu\over2\pi}{1\over\nu^2+1/4}e^{\omega(\nu)y}\eqno(8)
$$ 
 
\par\noindent 
with $\omega(\nu)\equiv\omega(\nu,0)$ as a consequence of the integration
on $\phi$. $\hat{\sigma}_L$ is the inclusive 
cross section for minijet production in a 
parton process represented 
by a cut BFKL Pomeron. A different possibility which one may 
consider is to produce two minijets without cutting the  
BFKL Pomeron[7], namely by elastic scattering of two partons 
which exchange a BFKL Pomeron at momentum transfer 
larger than the lower cut off $q_t^{min}$. 
The corresponding integrated partonic cross section is expressed 
as[3]: 
 
$$\hat{\sigma}_S=\Bigl({\alpha_sC_A\over\pi}\Bigr)^4 
  {\pi^5\over4(q_t^{min})^2}F_S(y)\eqno(9)$$ 
 
\par\noindent 
and $F_S(y)$ is the convolution of two BFKL propagators 
in the transverse momentum plane: 
 
$$F_S(y)={(q_t^{min})^2\over16\pi^3}\int_{ q_t^{min}}d^2k 
  \biggl{\vert}\int 
  d^2qd^2q'f^k(q,q',y)\biggr{\vert}^2\eqno(10)$$ 
 
\par\noindent 
$k$ is the overall momentum exchanged through the ladder and 
the label $S$ refers to the singlet exchange in the elastic 
partonic collision. Obviously 
one might consider also the possibility of having a 
octet exchange, the contribution to the cross section  
is however subleading at large rapidities[3]. 
The integrals on $q$ and $q'$ in Eq.(8) can be performed  
when using the integral representation of $f^k(q,q',y)$: 
 
$$\int d^2qd^2q'f^k(q,q',y)={4\over k^2}\int d\nu 
   {\nu^2\over(\nu^2+1/4)^2}{\rm 
exp}\bigl[\omega(\nu)y\bigr]\eqno(11)$$ 
 
\par\noindent 
and one obtains the asymptotic behavior at large $y$ 
 
$$\hat{\sigma}_S(y)\simeq\Bigl({\alpha_sC_A\over\pi}\Bigr)^4 
  {\pi^5\over4(q_t^{min})^2}\Biggl(\pi 
  {e^{4{\rm ln}2z}\over\bigl({7\over2}\zeta(3)\pi z\bigr)^{3/2}} 
  \Biggr)^2\eqno(12)$$ 
 
\par\noindent 
where $z=\alpha_sC_Ay/\pi$. For comparison the asymptotic 
behavior of $\hat{\sigma}_L$ is: 
 
$$\hat{\sigma}_L(y)\simeq\Bigl({\alpha_sC_A\over\pi}\Bigr)^2 
  {\pi^3\over2(q_t^{min})^2}\Biggl( 
  {e^{4{\rm ln}2z}\over\bigl({7\over2}\zeta(3)\pi z\bigr)^{1/2}} 
  \Biggr)\eqno(13)$$ 
 
\par 
The `elastic' partonic cross  
section can be identified with the `diffractive' cut of the double BFKL 
Pomeron  
exchange contribution to the forward parton amplitude[8]. At the same 
order, 
in the number of exchanged BFKL Pomerons,  
one needs therefore to keep into account also the 
one BFKL Pomeron cut and two BFKL Pomeron cut contributions to the 
partonic process. In fact the relative rate of elastic and inelastic 
parton processes, as expressed in Eq.(12) and 
in Eq.(13), is not consistent at large rapidity intervals because of
the too rapid rise of the elastic cross section. 
We do not try to solve here this unitarity problem. To have an
indication on the boundaries of the kinematical regime where the
problem needs to be faced we have taken the simplest attitude.
In analogy to the $s$-channel unitarization of the soft Pomeron exchange
we have included in the semihard partonic interaction the
exchange of two BFKL Pomerons and we have used the AGK 
cutting rules[9] to obtain the inelastic contributions to the  
cross section. The semihard partonic cross section  
$\hat{\sigma}_H(y)$ is therefore expressed as      
 
$$\hat{\sigma}_H(y)=\hat{\sigma}_S(y)+ 
  \bigl(\hat{\sigma}_L(y)-4\hat{\sigma}_S(y)\bigr)+ 
  2\hat{\sigma}_S(y) 
  \eqno(14)$$ 
  
\par\noindent 
where the single BFKL Pomeron exchange contributes with 
$\hat{\sigma}_L(y)$ and the contributions from the double 
BFKL Pomeron exchange, according with the AGK cutting rules,  
are: $\hat{\sigma}_S(y)$, the `diffractive' 
contribution i.e. neither Pomeron is cut, $-4\hat{\sigma}_S(y)$, the one Pomeron cut, and 
$+2\hat{\sigma}_S(y)$, the two Pomeron cut. Eq.(14) allows  
one to define the kinematical region of applicability of the approach. 
Indeed the one BFKL cut Pomeron contribution 
to the cross section must be positive[10]: 
 
$$\bigl(\hat{\sigma}_L(y)-4\hat{\sigma}_S(y)\bigr)>0\eqno(15)$$ 
 
\par\noindent 
Eq.(15) fixes a limiting value to the rapidity interval $y$ as a function 
of $\alpha_S$, as an example for $\alpha_s=.18$ and $\alpha_S=.22$ 
one obtains about $14$ and $10$ rapidity units respectively. 
Strictly speaking the contribution expressed by Eq.(15) implies
that the one-cut-Pomeron events and the two-cut-Pomeron events are
distinguishable in term of their multiplicity. If the fluctuations
in multiplicity prevents this identification we get the milder
condition

$$\bigl(\hat{\sigma}_L(y)-2\hat{\sigma}_S(y)\bigr)>0\eqno(15')$$ 
 
\par\noindent
which requires only the distinction between diffractive and 
non-diffractive events.
\par
We therefore express the semihard cross section $\hat{\sigma}_H(y)$,  
corresponding to a single partonic interaction, as 
 
$$\hat{\sigma}_H(y)=\hat{\sigma}_S(y)+ 
  \bigl(\hat{\sigma}_L(y)-2\hat{\sigma}_S(y)\bigr) 
  \equiv\hat{\sigma}_S(y)+\hat{\sigma}_P(y)\eqno(16)$$ 
 
\par\noindent 
and $\hat{\sigma}_P(y)$ 
is the contribution from production of secondaries, 
both from one and two cut BFKL Pomerons. 
 
\vskip.25in 
\par	{\bf 3. General Framework for Multiparton Interactions} 
\vskip.25in 
 
\par  
Given the scale $q_t^{min}$ one may separate  
parton fluctuations with a lifetime long with respect 
to $q_t^{min}$ and parton fluctuations with a lifetime short  
with respect to $q_t^{min}$. Parton fluctuations which have a long  
lifetime are of non-perturbative origin and can be associated to the  
initial state. Parton fluctuations with a short lifetime can be 
treated perturbatively and can be associated to the semihard 
interaction.   
In the simplest case, when the typical value of the rapidity interval 
$y$ in the partonic interaction is not too large, the semihard partonic 
interaction is described at the lowest order in the number of exchanged 
BFKL Pomerons. If $q_t^{min}$ is relatively small one faces however
a unitarity problem 
even if one is in a regime where the partonic interaction is 
well described within perturbation theory because of the large 
value of the integrated semihard hadronic cross section[11]: by integrating 
Eq.(5) with the cut off $q_t^{min}$ one obtains a cross section which  
easily exceeds the value of the total cross section.  
On the other hand the partonic cross section, 
Eq.(16), is in comparison still rather small. The large value  
of the integrated inclusive cross section is therefore the consequence 
of the large flux of partons in the initial state, which gives rise 
to an average number of partonic collisions larger than one[12]. 
In the typical 
semihard hadronic process different partonic interactions are 
localized in the transverse plane in different regions, of 
size of order $\langle\hat{\sigma}_H\rangle$ 
inside the overlap volume of the two interacting hadrons, whose 
transverse size is of a few $fm^2$. At Tevatron energy,  
considering partonic interactions with cross section $\hat{\sigma}_H$, 
as expressed in Eq.(16), and $q_t^{min}=6GeV$, the 
average distance in rapidity 
between produced minijets is about five units. The corresponding value 
of the partonic cross section is $\hat{\sigma}_H\bigl(\langle y\rangle\bigr)\simeq 
8\times10^{-2}mb$ and one may correspondingly estimate that 
in a central collision 
the average number of partonic interactions 
is $\simeq2\leftrightarrow 4$. These features characterize  
the kinematical regime  
which we are presently interested in: actually the `elementary' partonic 
interaction is well described within the BFKL dynamics,  
in such a way that the bound in Eq.(15) is satisfied, and 
the inclusive hadronic minijet cross section 
is larger than the inelastic cross  
section, in such a way that the rate of   
multiple parton interactions is sizeable.      
\par	To discuss multiple parton collisions 
we follow the approach used in ref.[6] and [13,14]:   
In the case of soft 
interactions  
multi-Reggeon exchanges are conveniently taken into account by making 
use of 
the AGK cutting rules[9]. Although no general  
proof of their validity is available in the case of 
semi-hard interactions, it has nevertheless been possible to show that 
the cutting rules hold for one of the components of the 
interaction which is leading in the large-$\hat{s}$ 
fixed-$\hat{t}$ limit[15]. If one assumes the validity of 
the cutting rules for semi-hard interactions, one is allowed to represent  
the semi-hard cross section $\sigma_H$ as a probabilistic distribution of  
multiple semi-hard parton collisions[14]. 
The most general expression for  
$\sigma_H$ requires however the introduction of 
the whole infinite set of multiparton distributions[16], which keep into 
account hadron fluctuations in the parton number. To that purpose  
we introduce the exclusive $k$-body parton distribution 
$W^{(k)}(u_1\dots u_k)$,  
namely the probabilities to find a hadron in a fluctuation with $k$ partons 
with 
coordinates $u_1\dots u_k$,  
 $u_i\equiv(b_i,x_i)$ standing for the transverse partonic coordinate 
$(b_i)$ 
and longitudinal fractional momentum $(x_i)$. From the exclusive 
distributions and the auxiliary functions $J(u)$ one may construct  
the generating functional ${\cal Z}[J]$: 
 
$${\cal Z}[J]=\sum_n{1\over n!}\int 
  J(u_1)\dots J(u_n)W_n(u_1,\dots u_n)du_1\dots du_n\eqno(17)$$  
 
\par\noindent 
in such a way that the exclusive distributions are the coefficients 
of the expansion of the generating functional around $J=0$, while the 
inclusive distributions are the coefficients of the expansion  
around $J=1$. A very general expression for the semihard cross section, 
which is consistent with the cutting rules, is therefore: 
 
$$\sigma_H=\int d^2\beta\sigma_H(\beta)$$ 
 
$$\eqalign{\sigma_H(\beta)=\int&\sum_n{1\over n!} 
  {\delta\over \delta J(u_1)}\dots  
  {\delta\over \delta J(u_n)}{\cal Z}_A[J]\cr 
  \times&\sum_m{1\over m!} 
  {\delta\over \delta I(u_1'-\beta)}\dots  
  {\delta\over \delta I(u_m'-\beta)}{\cal Z}_B[I]\cr 
\times&\Bigl\{1-\prod_{i=1}^n\prod_{j=1}^m\bigl[1-
\hat{\sigma}_H(u_i,u'_j)\bigr] 
   \Bigr\}\prod dudu'\Bigm|_{J=I=0}}  
\eqno(18)$$  
 
\noindent 
Here the  
$\beta$ is the impact parameter between the two interacting hadrons 
and $\hat{\sigma}_H(u_i,u'_j)$, 
 represents 
the probability for the parton $i$  
of the $A$-hadron to have a semihard interaction with the parton $j$  
of the $B$-hadron. 
The semi-hard cross section is constructed by summing over all possible 
partonic configurations of the two interacting hadrons (the sums over 
$n$ and $m$) and, for each configuration with $n$ $A$-partons and 
$m$ $B$-partons, summing over all possible multiple partonic 
interactions. This last sum is constructed by asking for the  
probability of no interaction between the two configurations 
(actually $\prod_{i=1}^n\prod_{j=1}^m[1-\hat{\sigma}_{i,j}]$ ). The 
difference from one of the probability of no interaction 
gives the sum over all  
semi-hard interactions. $\sigma_H(\beta)$ is then the probability to have 
at least one semihard parton interaction when the impact parameter in 
the hadronic collision is equal to $\beta$. The semi-hard cross section is 
obtained by integrating the probability $\sigma_H(\beta)$ on the impact 
parameter. 
Analogously, the elementary semi-hard cross section 
$\hat{\sigma}_H(x,x')$ is 
obtained by integrating the elementary interaction probability  
$\hat{\sigma}_H(u,u')$ 
on the relative transverse coordinate ${\bf b}-{\bf b}'$. 
\par	In Eq.(18) $\sigma_H$ is constructed by summing all possible 
semihard two-body parton collisions.  	 
Multiple semi-hard parton collisions are of two distinct kinds, 
disconnected 
collisions and rescatterings. In a disconnected collisions different pairs of 
partons  
interact independently at different points in the transverse plane. In a 
rescattering 
a high energy parton interacts several times, with momentum exchange 
larger 
than $q_t^{min}$, with different target partons and all interactions are 
localized 
in the same region, approximately of size $(1/q_t^{min})^2$, in the transverse 
plane. 
The most important contribution to the semi-hard  
cross section arises from the disconnected partonic collisions. 
In fact, at a given number of partonic collisions, the incoming
parton flux is maximized in the configuration where all collisions
are disconnected.
It might therefore be meaningful to obtain a simpler expression for $\sigma_H$ by 
neglecting the  
rescattering processes in Eq.(18). To that purpose we
expand the interaction probability 
( the factor in curly brackets ) as sums and suppress all addenda 
containing 
repeated indices: 
 
$$\Bigl\{1-\prod_{i,j}^{n,m}\bigl[1-\hat{\sigma}_{ij}\bigr]\Bigr\} 
\Rightarrow  
  \sum_{ij}\hat\sigma_{ij}-{1\over 2!} 
  \sum_{ij}\sum_{k\not=i,l\not=j}\hat\sigma_{ij}\hat\sigma_{kl}+ 
  \dots\eqno(19)$$  
 
\noindent 
Because of the symmetry of the derivative operators in Eq.(18) one can  
replace the expression in Eq.(19) with: 
 
$$nm\hat\sigma_{11}-{1\over 2!}n(n-1)m(m-
1)\hat\sigma_{11}\hat\sigma_{22} 
+\dots\eqno(20)$$ 
 
\noindent 
in such a way that the sums over $n$ and $m$ in Eq.(18) can be 
performed 
explicitly. 
As a consequence the cross section at fixed impact parameter 
$\sigma_H(\beta)$  
is expressed by the operatorial form: 
 
$$\sigma_H(\beta)= 
  \Bigl[ 1-{\rm exp}\Bigl(-{\delta\over\delta J}\cdot\hat{\sigma} 
  \cdot{\delta\over\delta I}\Bigr)\Bigr] 
  {\cal Z}_A[J+1]{\cal Z}_B[I+1]\Bigm|_{J=I=0} 
 \eqno(21)$$ 
 
\noindent 
where the dependence on the variables $u$ and $u'$ is understood.  
\par	The expression of $\sigma_H(\beta)$, as given by Eq.(21), is still too complicated to be 
worked out, since all possible multi-parton correlations are implicitly 
present in ${\cal Z}$. The simplest possibility is to neglect all 
correlations in the multi-parton distributions. In this case one writes 
 
$${\cal Z}[J+1]={\rm exp}\int D(u)J(u)du\eqno(22)$$ 
 
\noindent 
where $D(u)$ is the average number of partons. 
The cross section assumes therefore the eikonal form:  
 
$$\sigma_H=\int d^2\beta\Bigl[1-{\rm exp}\bigl(-\Phi(\beta)\bigr)\Bigr] 
  \eqno(23)$$ 
 
\par\noindent 
where 
 
$$\Phi(\beta)\equiv\Phi_S(\beta)+\Phi_P(\beta)\equiv 
  \int_{y_m}^{y_M}dy\int_{y}^{y_M}dy' 
  \bigl(\phi_S(\beta;y,y')+\phi_P(\beta;y,y')\bigr)\eqno(24)$$ 
 
\par\noindent 
with  
 
$$\phi_{S,P}(\beta;y,y')\equiv\int d^2bD_A\bigl(b,x(y)\bigr) 
  \hat{\sigma}_{S,P}(y'-y)D_B\bigl(b-\beta,x'(y')\bigr) 
  \eqno(25)$$ 
 
\par\noindent 
and $y_M$, $y_m$ are the maximum and minimum rapidity values 
allowed by kinematics. The index $S$ in 
Eq.(24) refers to the elastic parton interaction, the singlet exchange  
in Eq.(16), and the index $P$ to the inelastic interaction, the one 
and two cut Pomeron contributions in Eq.(16). The different contributions  
from multiple semihard parton collisions to the cross section are explicit  
if one expands the exponential in Eq.(23) as follows:

$$\eqalign{ 
  \sigma_H&=\int d^2\beta\sum_{\nu=1}^\infty{1\over\nu!}e^{-
\Phi(\beta)}\cr 
  &\times\int_{y_1<y_1'}dy_1dy_1' 
   \biggl(\phi_S(\beta;y_1,y_1')+\phi_P(\beta;y_1,y_1') 
   \biggr)\cr 
  &\dots\cr 
  &\times\int_{y_{\nu}<y_{\nu}'}dy_{\nu}dy_{\nu}' 
   \biggl(\phi_S(\beta;y_{\nu},y_{\nu}')+\phi_P(\beta;y_{\nu},y_{\nu}') 
   \biggr)} 
 \eqno(26)$$ 
 
\par	One is interested in the component of $\sigma_H$ 
which represents two minijets at rapidities 
$\bar{y}$ and $\bar{y}'$, in the central rapidity region, 
with associated gap $\Delta y=\bar{y}'-\bar{y}$  
in the rapidity distribution of secondary produced gluons. 
To that purpose one needs to exclude in Eq.(26) both the `elastic' terms, 
with final state minijets in the gap, and all the inelastic 
partonic interactions, generated with elementary probability 
$\phi_P$. In fact the cut BFKL Pomeron originates 
soft gluons distributed on the average uniformly in rapidity.  
Local fluctuations in  
the rapidity distribution of soft gluons, which could 
leave the gap empty also in the case of an inelastic elementary 
parton interaction, may be neglected 
if one considers a reasonably large rapidity gap. 
The cross section to observe two minijets at rapidities 
$\bar{y}$ and $\bar{y}'$, with the gap $\Delta y=\bar{y}'-\bar{y}$  
in the rapidity distribution 
of secondaries, is therefore expressed as 
 
$$\eqalign{ 
  {d\sigma_H(\Delta y)\over d\bar{y}d\bar{y}'}=\int d^2\beta 
  \Biggl[\sum_{\nu=1}^\infty\nu  
  \phi_S(\beta;\bar{y},\bar{y}') 
  {\bigl[\Phi_S(\beta,\Delta y)\bigr]^{\nu-1}\over\nu!}&\cr 
  +\sum_{\nu=2}^\infty\nu(\nu-1)  
  \int_{y_m}^{\bar{y}}dy\phi_S(\beta;y,\bar{y}') 
    \int_{\bar{y}'}^{y_M}&dy'\phi_S(\beta;\bar{y},y')\cr 
  \times&{\bigl[\Phi_S(\beta,\Delta y)\bigr]^{\nu-2}\over\nu!} 
  \Biggr]e^{-\Phi(\beta)}} 
  \eqno(27)$$ 
 
\par\noindent 
where 
 
$$\Phi_S(\beta,\Delta y)\equiv 
  \int_{y_m}^{\bar{y}}dy\int_{\bar{y}'}^{y_M}dy' 
  \phi_S(\beta;y,y')\eqno(28)$$ 
 
\par\noindent 
After summing on $\nu$ one obtains 
 
$$\eqalign{ 
  {d\sigma_H(\Delta y)\over d\bar{y}d\bar{y}'}=\int d^2\beta 
  \Bigl[\phi_S(\beta;\bar{y},\bar{y}') 
  +\int_{y_m}^{\bar{y}}&dy\phi_S(\beta;y,\bar{y}') 
    \int_{\bar{y}'}^{y_M}dy'\phi_S(\beta;\bar{y},y')\Bigr]\cr 
  \times&{\rm exp}\Bigl\{ 
   \Phi_S(\beta;\Delta y)-\Phi_S(\beta)-\Phi_P(\beta)\Bigr\} 
  }\eqno(29)$$ 
 
\par 
The two addenda in Eq.(29) are the single and double `elastic' 
scattering terms.  
In the single scattering term both observed minijets are 
produced in the same elementary partonic interaction, in the 
double scattering term the two minijets are generated in different partonic 
collisions. Both terms are multiplied by the absorption factor 
$exp\bigl\{-\bigl(\Phi_S(\beta)-\Phi_S(\beta;\Delta y)\bigr)\bigr\}$  
that removes the `elastic' parton interactions which would fill the gap,  
actually those which produce minijets with rapidities $y$ and $y'$ 
such that ${\bar y}\le y$ or $y'\le\bar{y}'$. 
At a fixed value of $\beta$ the cross section is multiplied 
by $exp\bigl\{-\Phi_P(\beta)\bigr\}$ which is the probability of not
having any inelastic partonic interaction in the process.
One may recognize in Eq.(29) the semihard contribution to the
survival probability factor $\langle S^2(\beta)\rangle$ of ref.[4]. Actually 
$exp\bigl\{-\Phi_S(\beta)-\Phi_P(\beta)\bigr)\bigr\}$ is the
probability factor of not having any semihard activity in the
underlying hadronic event.
A more detailed analysis of the origin of the survival factor and of the
suppression factor due to some elastic scattering is presented in the Appendix,
in the context of a multiparton dynamics. In the same Appendix the 
influence of the possible deviation from the Poissonian form of the original
parton distribution is also discussed.    
\vskip.25in 
\par	{\bf 4. Discussion and conclusions} 
\vskip.25in 
\par  
High energy hadronic interactions, with production of minijets and 
associated rapidity gap in the distribution of secondaries, is a 
process where one would expect to observe explicitly the unitarity relation 
at the level of partonic amplitudes. A closer inspection shows that  
the unitarity problem 
which one faces is twofold. On one hand, if one believes that the  
semihard interaction between partons can be described with the 
BFKL approach, one needs to solve the problem of the too rapid 
rise of the partonic cross section as a function of the rapidity 
interval between primary minijets. On the other hand, even if the  
size of the partonic cross section is not too large, namely if the 
hadronic c.m. energy is such that the typical rapidity interval  
between primary minijets is relatively small, the semihard hadronic      
cross section may still be too large. One can therefore distinguish 
three different regimes: 
\item{I-} The cutoff is sizeable with respect to the typical energy  
available to the semihard partonic interaction. The corresponding  
`elementary' parton interaction  
is small, no unitarization is needed and the semihard cross section  
is well described by a single partonic collision. 
\item{II-} The cutoff is moved towards relatively smaller values, with 
respect to the available energy. A single partonic interaction is 
still well described by the BFKL dynamics. The semihard 
hadronic cross section is however too large with respect to the 
total inelastic cross section and unitarity corrections are to be  
taken into account. The unitarization of the hadronic semihard  
cross section is achieved by taking into account multiparton 
interactions, namely different pairs of partons  
interacting independently with BFKL Pomeron exchange. 
Typically the different partonic interactions are localized at different  
points in the transverse plane, in the region of overlap  
of the matter distribution of the two hadrons. 
\item{III-} With even smaller values of the cutoff one may still be  
in the regime where perturbative QCD can be used, since the value of 
$\alpha_S(q_t^{min})$ is small, but the typical rapidity interval 
between primary minijets is too large so that the `elementary' 
parton process is not well described any more by the single 
BFKL Pomeron exchange and also the `elementary' parton 
process has to be unitarized. One may obtain an indication on the limits 
between  
regions II and III by testing whether the bound in Eq.(15) is 
satisfied. 
\par\noindent 
To have a quantitative indication on the boundaries  
of the kinematical regions,  
we have worked out a numerical example. Since the present available 
information on multiple parton interactions is limited to the 
scale factor which gives the rate of double parton interactions[17], 
while no information is available on the multiparton correlations, we 
have considered the simplest possibility, namely the Poisson 
distribution for the multiparton distributions. On the grounds that the main  
contribution to the multiple parton interactions 
is represented by the disconnected partonic collisions, we have 
neglected parton rescatterings. The unitarized expression for the 
cross section is therefore given explicitly as a function of the 
input which is used to evaluate the single scattering 
term, namely the average number of partons $D(b,x)$ and the `elementary'  
partonic cross section $\hat{\sigma}_H$. We have factorized $D(b,x)$ as 
$f_{eff}(x)\times F(b)$, where $f_{eff}(x)$ is the effective structure 
function as expressed in Eq.(6) and $F(b)$ is a Gaussian, normalized 
to one and such as to give for the double scattering term the scale factor 
$\sigma_{eff}$ (see Eq.(26) of Ref.[6]) 
consistent with the experimental indication[17].  
In our numerical example we have chosen $\sigma_{eff}=20mb$ and 
as a scale factor for the structure functions  
we have taken $q_t^{min}/2$. $\alpha_S$ is a free parameter in the BFKL 
approach, one expects however that the value of $\alpha_S$ which one 
should use is not too different from the value of 
the running $\alpha_S$ at the scale of the typical momentum transferred 
in the process. We have chosen as a value of $\alpha_S$ the value 
of the running coupling computed with $q_t^{min}/2$ as a scale factor. 
The values of the semihard cross section $\sigma_H$, as expressed in Eq.(23),
which we obtain with this input are consistent with the experimental values
published by UA1[18].  
At each value of the hadronic 
c.m. energy the boundary of the kinematical regions I, II and III
are identified by the choice of the cutoff $q_t^{min}$.
The curve which corresponds to larger values of $q_t^{min}$ in fig.1  
has been drawn requiring that the unitarized hadronic semihard cross  
section $\sigma_H$, as expressed in Eq.(23), is $20\%$ smaller with 
respect to the single scattering 
term. The lower curve corresponds to the value of $q_t^{min}$ which, 
on the average, namely after integration with the structure functions, 
saturate the bound in Eq.(15). The two curves identify the three regions 
mentioned above.  
\par	Moving from large values of $q_t$ to the semihard region one faces  
therefore two different unitarity problems, which signal
the appearance of different levels of structure in the hadronic  
interaction. The gap in the rapidity distribution of produced 
secondaries is an effect which derives from the unitarity 
relation applied to the `elementary' parton amplitude. The regime 
where `elementary' interactions with rapidity gap are a sizeable 
component of the `elementary' parton process is however a regime 
where the overall hadronic process is already structured 
in a non-trivial way, because of the large amount of multiparton 
interactions.
To have a quantitative feeling of the effect of multiparton interactions,
we have performed  
a numerical calculation in the simplest example 
already considered to obtain the curves in fig.1. The expression of
the cross section as a function of the rapidity interval $\Delta y$ 
is given in Eq.(29). We have considered $p\bar{p}$ interactions
at Tevatron energy and as a lower threshold to observe minijets
we have taken the value $q_t^{min}=5GeV$.
In fig.2 we plot the cross section, as expressed in Eq.(29), divided
by the survival probability factor $exp\bigl\{-\Phi_S(\beta)-
\Phi_P(\beta)\bigr\}$. The continuous curve is obtained by using as a
input the value $\sigma_{eff}=20mb$ and the dashed curve is obtained
by using $\sigma_{eff}=12mb$.
The dotted curve is the contribution of the single scattering term alone.
\par 
As it is shown in fig.2
the effect of unitarization on the behaviour of the 
cross section is large. 
In the actual case the main modification to the dependence on $\Delta y$
is due to the presence of multiple `elastic' parton scatterings whose effect
on the cross section is twofold. A different dependence on $\Delta y$, with
respect to the single scattering term, is
induced by the presence of two different sources. The first is
the contribution of the process where the two observed minijets 
originate in different `elastic' partonic interactions, 
the second term in Eq.(29),
the second source for the different dependence on $\Delta y$ is the correction
induced by multiple `elastic' scatterings to the survival probability 
factor. In fact
not all underlying hadron activity needs to be excluded. `Elastic' parton 
scatterings
which produce minijets outside the gap are allowed and the corresponding 
contribution to the cross section depends on $\Delta y$.
The effect of the inelastic semihard 
partonic interactions is, on the contrary, factorized 
at fixed impact parameter $\beta$ and independent on $\Delta y$.
The main effect of the inelastic partonic processes is to contribute  
to the survival probability $\langle S^2\rangle$ of ref.[4] 
rather than modifying the dependence on $\Delta y$. 
\par
The region 
where $q_t^{min}$ is relatively small is also the region which is closer to 
the BFKL kinematics and where, as a consequence, the BFKL approach to 
parton  
dynamics is better justified. In high energy hadronic interactions, 
at relatively small values of $q_t^{min}$, unitarity corrections are
however large and have to be taken into account. 
The physical effect which underlies the need of unitarization is the
increasing complexity of the semihard interaction which appears when
$q_t^{min}$ is lowered, as it is shown in fig.1. 
In the kinematical region which has been discussed in the present paper, 
corresponding with region II in fig.1, 
the `elementary' parton process is well described within the 
BFKL approach. The typical hadronic interaction is however 
characterized by several `elementary'  
partonic collisions, localized at different points in the 
transverse plane. One of the reasons of interest
in the actual kinematical regime is that, in region II, the  
non perturbative component of the process, 
which is factorized in the multiparton distributions, represents a new  
piece of information on the hadron structure with respect to the 
hadron structure functions of large $p_t$ physics. 
\vskip .25in
 {\bf Appendix}
\vskip 1 pc

{\it A.1 General features and inelastic scattering}
\vskip 1pc
In this Appendix the way in which an overall term independent of $\Delta y$
and further corrections explicitly dependent on  $\Delta y$ may arise out
of the many parton dynamics is discussed more in detail.
The functional formulation, which has been already employed, is used again
and more systematically. The longitudinal momenta are always given in terms
of the fractional momentum $x$, the relation with the corresponding rapidity
$y$ was mentioned in Sec.2. 
\par
If we look to configurations where there are two radiated partons (jets) with 
rapidity gap sitting in the central rapidity region, in the two-hadron c.m.,
then the two partons belong to different original hadrons. The parton
population is divided into three segments $x_a,x_b,x_c$: the $x_a$ interval 
correspond to the partons which would fill the gap, so to the
parton that, in the configuration looked for, are not scattered; the $x_b$
correspond to the partons outside the gap, so to the partons which are simply
not observed; the $x_c$ correspond to the parton observed at one end of the 
gap; the segment in which $x_c$ lies will be ideally shrunk to one point. 
The three regions of variations of $x$ reflect into three regions of variations
of $u$, but clearly the impact parameter {\bf b} is not affected by this 
operation.
\par
This decomposition suggests a way of rewriting the generating functional which
is sometimes more convenient :
$$\eqalign{ {\cal Z}[J]&=
\sum_n {1 \over n!} \int J(u_1)\cdots  J(u_n) W_n(u_1,...u_n) 
  du_1 ...du_n \cr
&=\sum_{p,q,k} {1 \over p!} {1 \over q!} {1 \over k!} 
\int J(u_a)\cdots  J(u_b)\cdots  J(u_c) W_{p,q,k}(u_a,...u_b,...u_c) 
 du_a ... du_b ...du_c,\cr} \eqno (A.1) $$
$p$ is the number of partons of kind $a$, $q$ is the number of partons of 
kind $b$, $k$ is the number of partons of kind $c$, and an analogous 
expression is written for ${\cal Z}[I(u')]$.
\par
We start considering the effect of inelastic processes and 
we express the probability that:
\par
I- A parton $u_c$ scatters elastically against a parton $u'_c$ or a parton 
$u_c$ scatters elastically against a parton $u'_b$ and parton $u'_c$ scatters 
elastically against a parton 
$u_b$ thus giving rise to the two partons at the end of the gap and nothing is
produced in the middle.
\par
II- No inelastic scattering takes place, since such a process would give rise
also to partons which could fill the gap.
\par
In formulae the requirement I is expressed by an operator ${\cal F}$ while the 
requirement II is expressed by an operator ${\cal G}$, both applied to the 
product  ${\cal Z}[J]{\cal Z}[I]$.
\par
The actual form of the two operators is:

$$\eqalign{ 
 {\cal F}=&\int du_c du'_c\Bigl[ \delta_c(u) 
   \hat\sigma_E (u_c,u'_c)  \delta '_c(u') +\cr 
             &\Bigl(\int \delta_c(u) \hat\sigma_E (u_c,u'_b) \delta '_b(u') \;
                         \delta_b(u) \hat\sigma_E (u_b,u'_c) \delta '_c(u')
   du_b du'_b \Bigr) \Bigr]\cr
  } \eqno (A.2)$$

$$\eqalign{ 
 {\cal G}=\sum_{n,n'}& \int \int  
 {1 \over n!} {1 \over n'!}
 \delta_1(u)\cdots \delta_n(u)\cdots \delta '_1(u') \cdots \delta '_{n'}(u')
 \times\cr
 &\prod  [1-\hat\sigma_{In} (u,u')] du du' \cr
} \eqno (A.3)$$

 In the definition of ${\cal G}$ there is no point in distinguishing the
 intervals of rapidity.
 Some notational simplification is obtained by setting:
 $$
    {\delta \over {\delta J(u_m)}}=\delta_m (u)\qquad
      {\delta \over {\delta I(u'_m-\beta)}}=\delta '_m (u')\qquad m=a,b.
 \eqno (A.4 ) $$

\vskip 1pc
 A first step in the choice of some specific distribution could be to neglect 
 the genuine many-body correlations, in this
 case the generating functional is in fact reduced to an ordinary function of a
 linear functional of the sources
 $${\cal Z}[J]=\Phi({\cal Y})\qquad\qquad {\cal Y}=\int D(u) J(u) du \eqno(A.5)
 $$
 The division of the field of variation of $x$ into three parts induces a
 corresponding decomposition ${\cal Y}={\cal Y}_a+{\cal Y}_b+{\cal Y}_c$.
 \par 
 Even with the restriction to a pure one-body density the general expression 
 is complicated. A very relevant and effective
 simplification is produced if we assume a Poissonian distribution for the
 initial partonic distributions
   $${\cal Z}[J]=\exp [{\cal Y}-{\cal Y}_o] \eqno(A.6).$$
 The normalization term ${\cal Y}_o=\int D(u) du$ ensures that ${\cal Z}[1]=1$.
\par
 In looking for some general features of the inelastic rescattering it is 
 however possible to use a general partonic distribution without many-body
 correlations, as it is described by the generating functional $\Phi ({\cal Y})$
 ; so we let the functional differential operators ${\cal G}$ and ${\cal F}$
 act on the  product  $\Phi ({\cal Y}[J]) \Phi ({\cal Y}[I])$
 at the end the auxiliary sources are put to zero: $J=I=0.$
 From the action of the operator ${\cal G}$ one obtains the intermediate 
 result:

$$\eqalign{ 
 {\cal K}[J,I]=\sum_{n,n'}& \int \int  
 {1 \over n!} {1 \over n'!}
 D(u_1)\cdots D(u_n)\cdots
 D(u'_1-\beta)\cdots D(u'_{n'}-\beta)\times \cr
 &\prod  [1-\hat\sigma_{In} (u,u')] du du' 
 \Phi^{(n)} ({\cal Y}[J]) \Phi^{(n')} ({\cal Y}[I]) \cr
} \eqno (A.7)$$

 From the further action of the operator ${\cal F}$ and the condition  $J=I=0$
 one obtains the final result:

$$\eqalign{ 
 K(\beta)=&\sum_{n,n'} \int \int  
 {1 \over n!} {1 \over n'!}
 D(u_1)\cdots D(u_n)\cdots D(u'_1-\beta)\cdots D(u'_{n'}-\beta)\cr 
 & \prod  [1-\hat\sigma_{In} (u,u')] du du' \times \cr
 &\int du_c du'_c \Bigl[D(u_c)  \hat\sigma_E (u_c,u'_c) D(u'_c-\beta)
 \Phi^{(n+1)} (0) \Phi^{(n'+1)} (0)+ \cr
  &\int D(u_c) \hat\sigma_E (u_c,u'_b) D(u'_b-\beta)
  \,  D(u_b) \hat\sigma_E (u_b,u'_c) D(u'_c-\beta)  du_b du'_b\cr 
 &\Phi^{(n+2)} (0) \Phi^{(n'+2)} (0)\Bigr]\cr
} \eqno (A.8)$$
So in general the expression has the form

$$ K(\beta)=\sum_{n,n'}[ G_{n,n'}^I(\beta)\cdot F^I(\beta)+ 
 G_{n,n'}^{II}(\beta)\cdot F^{II}(\beta)] \eqno (A.9)$$

For a Poissonian partonic distribution it results for every derivative
   $\Phi^{(n)}(0)=\exp [-{\cal Y}_o] $, so the double sum over
$n,n'$ can in principle be carried out yielding an overall
factor $G(\beta)=\sum_{n,n'}[ G_{n,n'}^I(\beta)+ G_{n,n'}^{II}(\beta)]$ 
which multiplies the remaining term

$$\eqalign{ F(\beta)=
& \int du_c du'_c \Bigl[D(u_c)  \hat\sigma_E (u_c,u'_c) D(u'_c-\beta) +\cr
  &\int D(u_c) \hat\sigma_E (u_c,u'_b) D(u'_b-\beta)
  \,  D(u_b) \hat\sigma_E (u_b,u'_c) D(u'_c-\beta)  du_b du'_b\Bigr].}
 \eqno (A.10)$$
 In other words it results that, at fixed hadronic impact parameter $\beta$
 the inelastic processes give simply rise to a multiplicative factor
 to the fundamental amplitude.
\par
In presence of another kind of distribution the treatment is less 
straightforward because the sum over $n,n'$ does not allows the extraction
of a common term $ F(\beta)$. It seems however likely that
in the expression of $K$ the second addendum, containing $F^{II}$ is more 
important than the first one, because it involves a less exclusive condition, 
if this is true then the sum
$$\bar G(\beta)=\sum_{n,n'} G_{n,n'}^{II}(\beta)$$
can be still be carried out yielding again a multiplicative factor in front of
the fundamental amplitude at fixed hadronic impact parameter.
\par
This essential simplicity is destroyed if we are in presence of sizeable 
two-body correlations. To be definite we may consider an example where we 
have a Poissonian distribution corrected by a two-body correlation
   $${\cal Z}[J]=\exp [{\cal Y}-{\cal Y}_o].$$
   $${\cal Y}=\int D(u)J(u) du+\m \int C(u,\bar u)J(u) J(\bar u) du d\bar u. 
   \eqno (A.11)$$
Then in performing the derivatives in order to calculate $K(\beta )$ we end 
unavoidably with expressions where the correlation term $ C(u,\bar u)$ links 
the inelastic component with
the elastic fundamental one and so the decomposition given in Eq.(A.10) is no
longer possible, in this case we expect that the inelastic processes modify
strongly not only the size but also the shape of the fundamental elastic
process.

\vskip 1 pc
 {\it A.2 Effect of the multiple elastic scattering }
\vskip 1pc
It is clear that in presence of a large flux of partons the inelastic processes
that have been considered till now are not the only disturbing effects $i.e.$
there are other dynamical processes which also act in the sense of filling the
rapidity gap that the "fundamental" process described in the present formalism 
by ${\cal F}$ produces. With reference to the notation introduced in eq (A.1),
one must take into account also the following elastic processes:
\par
A parton $u_a$ which scatters against a parton $u'_a$, this event will fill the
gap. A parton  $u_a$ which scatters against a parton $u'_b$ or a parton $u_b$
 which scatters against a parton $u'_a$, this event will also fill the gap.
The statement that these kind of collisions must not happen is expressed by a
$veto$ operator which, in absence of inelastic interactions would have the 
following form:

$$\eqalign{ 
 {\cal E}=\sum_{p,p'} \sum_{q,q'}& \int \int  
 {1 \over p!} {1 \over q!} {1 \over p'!} {1 \over q'!}
 \delta_a(u)\cdots \delta_b(u)\cdots
 \delta '_a(u') \cdots \delta '_b(u')\times \cr
 &\prod  [1-\hat\sigma_E(u_a,u'_a)]\cdot [1-\hat\sigma_E(u_a,u'_b)]\cdot 
 [1-\hat\sigma_E(u_b,u'_a)] du_a du_b du'_a du'_b \cr
} \eqno (A.12)$$

 The fact that the parton of kind $b$ are unobserved suggests that the sum
 over $q,q'$ is the first operation to be performed.

For a fixed $p$ the term containing the sum over $q'$ is
$$ \eqalign{
   \Delta (u_{a,1},\cdots,u_{a,p})=& \delta_a(u_1)\cdots \delta_a(u_p)\times \cr
   \sum_{q'}{1\over{q'!}}\prod_k &\Bigl[\int[1-\hat\sigma_E(u_{a,1},u'_k)]
   \cdots [1-\hat\sigma_E(u_{a,p},u'_k)]\delta'_b(u'_k) du'_k\Bigr].\cr
} \eqno (A.13)$$
Since all the $u'_k$ are dummy variables the sum over $q'$ can be carried out
and gives an exponential; then the product of the $[1-\hat\sigma]$ factors in 
the exponent may be expanded in terms containing no $\hat\sigma$, one 
$\hat\sigma$, two $\hat\sigma$, and so on.
$$ \eqalign{
  \Delta &(u_{a,1},\cdots,u_{a,p})= \delta_a(u_1)\cdots \delta_a(u_p)\times \cr
   &\exp \Bigl[\int \Bigl(1-\sum_j\hat\sigma_E(u_{a,j},u'_k)+
   \sum_{i\neq j} \m \hat\sigma_E(u_{a,i},u'_k) \hat\sigma_E(u_{a,j},u'_k)+ \cdots
   \Bigr) \delta'_b(u'_k) du'_k\Bigr],\cr
} \eqno (A.14)$$
The sum over $q$ is obviously treated in the same way and we get for ${\cal E}$
the expression:
$$ 
 {\cal E}=\sum_{p,p'} \int  
 {1 \over p!} {1 \over p'!}
 \prod  [1-\hat\sigma_E(u_a,u'_a)] 
  \Delta (u_{a,1},\cdots,u_{a,p}) \Delta (u'_{a,1},\cdots,u'_{a,p'})
\prod du_adu'_a \eqno (A.15)
$$
In order to obtain more explicit expressions it is necessary, now, to expand
both the operators $\Delta$ and the term $\prod [1-\hat\sigma]$ in multiple 
interactions, but at this step it seems that the treatment becomes clearer if
one chooses some definite form of the parton distributions.

\vskip 1 pc
 We choose now a Poissonian distribution of partons 
\footnote*{In order to perform the actual calculation it may be useful to 
 remember the identity
 $f(d/dx)g(x)|_{x=0}=g(d/dx)f(x)|_{x=0}$, which is evident whenever the two
 functions admit a power expansion around 0; in particular it yields
 $f(d/dx)\e^{hx}|_{x=0}=f(h)$} 
and let the functional differential operator ${\cal E}$ act on the  product 
  ${\cal Z}[J]{\cal Z}[I]$, after the action of  ${\cal F}$.
 Only at the end the auxiliary sources are put to zero:
 $J=I=0$. As it was already said, for the operator ${\cal E}$ one is forced to
proceed in steps corresponding to successive reinteractions of the same partons:
 from the operator $\Delta$ one gets:
$$ \eqalign{
   A&(u_{a,1},\cdots,u_{a,p})=D(u_{a,1})\cdots D(u_{a,p})\times \cr
   &\exp \Bigl[\int [1-\sum_i\hat\sigma_E(u_{a,i},u'_b)+
   \sum_{i\neq j} \m \hat\sigma_E(u_{a,i},u'_b) \hat\sigma_E(u_{a,j},u'_b)+ \cdots]
   D(u'_b-\beta) du'_b\Bigr].\cr
} \eqno (A.16)$$
 Also the term $P=\prod [1-\hat\sigma_E(u_{a,i},u'_{a,r})]\equiv 
 \prod [1-\hat\sigma_{i,r}]$ is expanded with the result:
  $$
   P=1-\sum \hat\sigma_{i,r}+\sum\sum \m
  \hat\sigma_{i,r}\hat\sigma_{j,s}-\sum\sum\sum (1/3!)  
 \hat\sigma_{i,r}\hat\sigma_{j,s}\hat\sigma_{k,t}+\cdots.
  \eqno (A.17)$$
 In the repeated sums the pairs of indices cannot be equal $e.g.\;(i,r)\neq
 (j,s)$; it is however possible to have either $i=j$ or $r=s$, what correspond
 to a rescattering of a particular parton.
 If we took only the term 1 and the simple sum in the exponent of $\Delta$,
 out of the operator ${\cal E}$ we would get this result:
 $$\eqalign{
E_o=&\exp \Bigl[\int D(u_a)\Gamma (u_a-\beta)du_a \Bigr]
     \exp \Bigl[\int D(u'_a-\beta)\Gamma (u'_a) du'_a \Bigr]\times \cr
    &\exp \Bigl[\int D(u_b)du_b+\int D(u'_b)du'_b\Bigr]\,. \cr}
\eqno (A.18)$$
 In this formula the absorption factors $\Gamma$ have been introduced, in their 
 definition the property that $\hat\sigma$ depends only on the difference
 of the impact parameters has been used.
 $$\Gamma (u_a-\beta)=\exp \Bigl[-\int \hat\sigma_E (u_a,u'_b)D(u'_b-\beta)du'_b
 \Bigr]$$
 $$\Gamma (u'_a)=\exp \Bigl[-\int \hat\sigma_E (u'_a,u_b)D(u_b)du'_b \Bigr].$$
 
 The factor $E_o$ takes into account the interaction of the partons
 of kind $a$ with those of kind $b$, but not the interaction of the $a$-partons
 among themselves.
 It could be put in a form that could be of easier interpretation. $i.e.$

 $$\eqalign{
E_o=
    &\exp \Bigl[\int D(u)du+\int D(u')du'\Bigr]\times \cr
    &\exp \Bigl[-\int D(u_a)[1-\Gamma (u_a-\beta)]du_a \Bigr]
     \exp \Bigl[-\int D(u'_a-\beta)[1-\Gamma (u'_a)]du'_a \Bigr]\;. \cr}
\eqno (A.18')$$
 \par
 In this expression the first exponential, where also the small contribution
 from $u_c$ has been included, takes simply away the normalization term 
  $\exp [-{\cal Y}_o-{\cal Y}'_o]$, it does not belong to the multiple
 scattering, while the second term, $E'_o$ is an "absorption"
 term due to some of the disconnected collision which fill the gap.
 We can do better and consider the other terms in the expansion of
 $P= \prod [1-\hat\sigma_{i,j}]$, with the restriction of excluding the repeated
 interaction of the same $a$-parton, in formulae $i\neq j\;and\;r\neq s$ in 
 this way a further factor is produced which multiply the former expression,
$i.e.$
$$
  E_1= \exp \Bigl[-\int D(u_a)\Gamma (u_a-\beta)
 \hat\sigma_E (u_a,u'_a)D(u'_a-\beta) \Gamma (u'_a)du_adu'_a\Bigr]
 \eqno (A.19)
  $$
 Going on means to produce expressions quite elaborated, one could, as an 
 example, calculate the factors which describe a double interaction of 
 partons either of kind $a$ or of kind $b$. When the partons $a$ are allowed
 to collide twice, but not three times, the following factor is produced:
$$\eqalign {
  E_2=
\exp \m \Bigl[\int &D(u_a)\Gamma (u_a-\beta)\hat\sigma_E(u_a,u'_a)D(u'_a-\beta)
\times \cr
&\Gamma(u'_a)\hat\sigma_E(u'_a,u_{\bar a})D(u_{\bar a})
  \Gamma (u_{\bar a}-\beta)du_adu_{\bar a}du'_a +\cr   
\int &D(u'_a-\beta)\Gamma (u'_a)\hat\sigma_E(u'_a,u_a)D(u_a)\times \cr
& \Gamma(u_a-\beta)\hat\sigma_E(u_a,u'_{\bar a})D(u'_{\bar a}-\beta)
  \Gamma (u'_{\bar a})du_adu'_adu'_{\bar a}\Bigr].\cr
 }\eqno (A.20)$$
 The expansion can be further continued, $e.g.$ by considering also the 
 collision of two partons of kind $b$ and then the various results can be 
  collected. From the term explicitly calculated one gets:
 $E \approx E'_oE_1E_2\cdots$.
 This factor modifies qualitatively the result expressed in Eq.(A.9): 
 because in
 that expression the terms $G_{n,n'}$ were independent of the rapidities
 defining the gap which appear in  $F_{n,n'}$ through $u_c=(x_c,{\bf b})$;
 now, on the contrary these variables appear explicitly in the terms $E $. 

\par
 Other form of partonic distributions give rise to more complicated final 
 expressions, but some of them allow anyhow some elaboration, in particular
 as already remarked in [14], the generating functional of a negative binomial 
 distribution allow a representation which is an integral transform of a Poisson
 generating functional:
$$
 \Phi({\cal Y})= {(1-{\cal Y}_o)^{\alpha} \over (1-{\cal Y})^{\alpha}}=
{1\over {\Gamma (\alpha)}} (1-{\cal Y}_o)^{\alpha} \int
             \e^{t{\cal Y}} \e^{-t} t^{\alpha -1} dt \eqno(A.21)
$$
 So, in principle, from the previous expressions holding for a Poisson
 distribution of partons it would be possible to get the analogous one for a 
 negative binomial distribution by applying the following prescriptions:
\par
I-Take away the normalization term, 
${\cal N}_P=\exp[-{\cal Y}_o-{\cal Y}_o']$
 and substitute it by
\par
 ${\cal N}_B=[\Gamma (\alpha)]^{-2} (1-{\cal Y}_o)^{\alpha} 
 (1-{\cal Y}_o')^{\alpha}$
\par
II-Multiply by a factor $t$ every $D(u)$ explicit or implicit, $i.e.$ inside 
$\Gamma $ and $B$;
\par
and multiply by a factor $s$ every $D(u')$ explicit or implicit, $i.e.$ inside
$\Gamma $ and $B$.
\par
III-Multiply the overall resulting expression by $\e^{-t-s}(ts)^{\alpha -1}$.
\par
IV-Integrate the result in $dt\,ds$ from 0 to $+\infty$.
\par
It is evident that the Poissonian and the negative binomial distribution for the
original partons are only the simplest ones which can be chosen; in term of 
integral transforms of the generating functions other distributions could be
built up. 
\vskip 1pc

{\it A.3 Elastic and inelastic scattering}
\vskip 1pc
The real case implies evidently both elastic and inelastic scattering. The
requirement that there be no production process filling the gap is expressed
by an operator of the form

$$\eqalign{ 
 {\cal C}=\sum_{p,p'} \sum_{q,q'}& \int \int  
 {1 \over p!} {1 \over q!} {1 \over p'!} {1 \over q'!}
 \delta_a(u)\cdots \delta_b(u)\cdots
 \delta '_a(u') \cdots \delta '_b(u')\times \cr
 &\prod  [1-\hat\sigma_T(u_a,u'_a)]\cdot [1-\hat\sigma_T(u_a,u'_b)]\cdot 
 [1-\hat\sigma_T(u_b,u'_a)] \cr
 &\prod [1-\hat\sigma_{In}(u_b,u'_b)]du_a du_b du'_a du'_b \cr
} \eqno (A.22)$$

$$\hat\sigma_T=\hat\sigma_E+\hat\sigma_{In} \eqno (A.23)$$
The statement expressed by this operator is that while the parton of kind 
$a$ must not suffer interactions at all, the partons of kind $b$ may
interact provided their interaction is elastic.
In its full form this expression appears quite intractable, however one can 
follow the idea that the most relevant dynamical feature is the inelastic 
scattering and the elastic is a perturbation. It does not seems 
convenient, however, to start from the inelastic operator ${\cal G}$, it is
more efficient to start with an "exaggerated" form of the operator ${\cal C}$ 
such that also the elastic interactions of the partons $b$ are excluded, 
by defining

$$
 {\cal C}_o=\sum_{n,n'} \int \int  
 {1 \over n!} {1 \over n'!}
 \delta_1(u)\cdots \delta_n(u)\cdots \delta '_1(u') \cdots \delta '_{n'}(u')
 \prod  [1-\hat\sigma_T (u,u')] du du' \;, \eqno (A.24)$$
 which looks very similar to ${\cal G}$, eq(A.4) and so yields, by itself, a
 factor independent of $y$.
 Then one has to correct the result by an expansion in $\hat\sigma_E $ that
 uses eq (A.23).
 Since the coordinates $u$ are all equivalent the expansion has the same
 combinatorial structure as:
  $$[A+B]^{qq'}=A^{qq'}+qq' B A^{(q-1)(q'-1)}\cdot A^{q-1}\cdot A^{q'-1}+
 B^2\cdots $$
 In this way, through a shift in $q$ and $q'$, the integrand of the operator
  ${\cal C}_o$ is reproduced with additional factors. The actual form of the
 first order correction in  $\hat\sigma_E $ is:
 
$$\eqalign{ 
 {\cal C}_1=&\sum_{p,p'} \sum_{q,q'} \int \int  
 {1 \over p!} {1 \over q!} {1 \over p'!} {1 \over q'!}
 \delta_a(u)\cdots \delta_b(u)\cdots
 \delta '_a(u') \cdots \delta '_b(u') \delta _b(v) \delta '_b(v')\times \cr
 &\prod  [1-\hat\sigma_T(u_a,u'_a)]\cdot [1-\hat\sigma_T(u_a,u'_b)]\cdot 
 [1-\hat\sigma_T(u_b,u'_a)]\cdot [1-\hat\sigma_T(u_b,u'_b)]\times\cr
 & du_a du_b du'_a du'_b \cr
 &\int \hat\sigma_E(v_b,v'_b)\prod [1-\hat\sigma_T(u_b,v'_b)]
\prod [1-\hat\sigma_T(v_b,u'_b)] \delta_b(v) \delta '_b(v')
  dv_b dv'_b\;.
} \eqno (A.25)$$
 In a limit in which terms like $\hat\sigma_E \cdot \hat\sigma_T$ are
 neglected, while keeping the powers of  $\hat\sigma_T$ the correction may 
 be factorized in the form
 $$ {\cal C}_1\approx  {\cal C}_o\cdot 
 \int \hat\sigma_E(v_b,v'_b) \delta_b(v) \delta '_b(v')
  dv_b dv'_b\;.\eqno (A.26')$$
 This approximation for the correction $ {\cal C}_1 $ suggests also a 
 generalization. The expression itself originates from the division of the whole
 flux of parton in a pair suffering only elastic scattering and in a rest for
 which all interaction are considered. This attitude can be extended to more
 general situation, on the ground that, 
as already noticed, in presence of large parton fluxes
 the multiple disconnected interactions may become more relevant than the 
 reinteractions. So, with reference to Eq.(A.22) we can divide the $q$ partons
 of kind $b$ into two subsets of $r$ and $s$ partons, obviously with $r+s=q$,
 and the same for $q'$. The coordinates of the $r$ partons will be still denoted
 by $u$, the coordinates of the $s$ partons will be called $v$.
 In this way Eq.(A.22) takes the following form:
 
$$
 {\cal C}=\sum_{p,p'} \sum_{r,r'} \sum_{s,s'} \int 
 {1 \over p!}{1 \over r!}{1 \over s!} {1 \over p'!}{1 \over r'!}{1 \over s'!}
 \delta_a(u)\cdots  \delta '_b(v')\cdot
 \prod  [1-\hat\sigma]du_a du_b dv_b du'_a du'_b dv'_b\;,  \eqno (A.27)$$
 
 the product $\prod  [1-\hat\sigma]$ contains nine kinds of factors.
 Now we decide, on the basis of the physical consideration outlined before
 that the $v$ parton suffer only elastic scattering, without any other 
 interaction, then necessarily $s=s'$, and there are $s!$ ways of pairing the
 $v$ with the $v'$. So it is not difficult to see that the rest of ${\cal C}$
 reconstruct the a factor ${\cal C}_o$ of Eq.(15), whereas the sum over $s$
 yields an exponential. In this way the approximate expression for ${\cal C}$
 is given by:
   $${\cal C} \approx {\cal C}_o\cdot \exp \Bigl[
  \int  \delta_b(v') \hat\sigma_E(v_b,v'_b)  \delta '_b(v') dv_b dv'_b\Bigr]\;.
  \eqno (A.27')$$

 When the parton distribution is Poissonian out of Eq.(A.27') one gets
 the expression of the first correction due to multiple scattering depending
 on $\Delta y$ which corresponds to what was previously stated in section 3,
 Eq.(29). 
 Anyhow, in a way wholly independent of the validity of the approximation we
 see that the correction is always positive because it represents a 
 compensation for the previous 
 exclusion of configurations that should be allowed, the amount of the 
 correction diminishes by enlarging the rapidity gap since the integration 
 runs over the complementary domain.
 For our purposes the most relevant feature of the elastic interactions is 
 that they produce effects depending from the very beginning on
 the $y$ of the gap because the integrations over $v,v'$ depend explicitly
 on $x_c\,,x'_c$ which are precisely the variables that set the difference
 between partons of kind $a$ and partons of kind $b$.

\vfill 
\eject

\par    {\bf References} 
\vskip .15in 
\item{1.} B.M. McCoy and T.T. Wu, {\it Phys. Rev. Lett.} {\bf 35}, 604 
(1975); 
{\it Phys. Rev.} {\bf D13}, 1076 (1976); V.S. Fadin, E.A. Kuraev 
and L.N. Lipatov, {\it Phys. Lett.} {\bf B60}, 50 (1975); L.{\L}ukaszuk 
and 
X.Y. Pham, {\it Phys. Lett.} {\bf B53}, 287 (1974); H.T. Nieh and Y.P. 
Yao, 
{\it Phys. Rev.} {\bf D13}, 1082 (1976); L. Tyburski, {\it Phys. Rev.} 
{\bf D13}, 1107 (1976); L.L. Frankfurt and V.E. Sherman, {\it Sov. J. 
Nucl. 
Phys.} {\bf 23}, 581 (1976); C.Y. Lo and H. Cheng, {\it Phys. Rev.} 
{\bf D13}, 1131 (1976); P.S. Yeung, {\it Phys. Rev.} {\bf D13}, 2306 
(1976); 
H. Cheng and C.Y. Lo, {\it Phys. Rev.} {\bf D15}, 2959 (1977); 
H. Cheng and T.T. Wu, {\it Expanding Protons: Scattering at 
High Energies}, Cambridge, Mass., The MIT Press, 1987; 
H. Cheng, J.A. Dickinson and K. Olaussen, {\it Phys. Rev.} 
{\bf D23}, 534 (1981). 
\item{2.} M.T. Grisaru, H.J. Schnitzer and H.S. Tsao {\it Phys. Rev. Lett} 
{\bf 30}, 811 (1973); L.N. Lipatov, {\it Yad. Fiz.} {\bf 23}, 642 (1976); 
E.A. Kuraev, L.N. Lipatov and V.S. Fadin, {\it Zh. Eksp. Teor. Fiz.} 
{\bf 71}, 840 (1976) [{\it Sov. Phys. JEPT} {\bf 44}, 443 (1976)]; 
{\bf 72}, 377 (1977) [{\it Sov. Phys. JEPT} {\bf 45}, 199 (1977)]; 
Ya.Ya. Balitskii and L.N. Lipatov, {\it Yad. Fiz.} {\bf 28}, 
1597 (1978) [{\it Sov. J. Nucl. Phys.} {\bf 28}, 822 (1978)]; 
{\it Pis'ma Zh. Eksp. Teor. Fiz.} {\bf 30}, 383 (1979) 
[{\it JETP Lett.} {\bf 30}, 355 (1979)];
V. Del Duca, preprint DESY 95-023, DFTT 13/95 (unpublished);
V.T. Kim and G.B. Pivovarov, {\it Phys. Rev.} {\bf D53}, 6 (1996);
{\it Phys. Rev.} {\bf D54}, 725 (1996); M.A. Braun, {\it Phys. Lett.}
{\bf B345}, 155 (1995); {\it Z. Phys.} {\bf C70}, 103 (1996);
N.N. Nikolaev, G.B. Zakharov and V.R. Zoller, {\it Phys. Lett.}
{\bf B366}, 337 (1966).    
\item{3.} V. Del Duca and W.-K. Tang, {\it Phys. Lett.} {\bf B312}, 225 
(1993). 
\item{4.} J.D. Bjorken, {\it Phys. Rev.} {\bf D47}, 101 (1992). 
\item{5.} E. Gotsman, E.M. Levin and U. Maor, {\it Phys. Lett.} {\bf 
B353}, 
526 (1995). 
\item{6.} R. Ragazzon and D. Treleani, {\it Phys. Rev.} {\bf D53}, 55 
(1996). 
\item{7.} A.H. Mueller and W.-K. Tang, {\it Phys. Lett.} {\bf B284}, 123 
(1992). 
\item{8.} L.V. Gribov, E.M. Levin and M.G. Ryskin, {\it Phys. Rep.} 
{\bf 100}, 1 (1983). 
\item{9.} V. Abramovskii, V.N. Gribov and O.V. Kancheli, {\it Yad. Fiz.} 
{\bf 18}, 595 (1973) [{\it Sov. J. Nucl. Phys.} {\bf 18}, 308 (1974) ]. 
\item{10.} L. Frankfurt, W. Koepf and M. Strikman {\it Phys. Rev.}
{\bf D54}, 3194 (1996).
\item{11.} G. Pancheri and Y. Srivastava {\it Phys. Lett.} {\bf B182}, 
199 (1986); S. Lomatch, F.I. Olness and J.C. Collins {\it Nucl. Phys.} 
{\bf B317}, 617 (1989).  
\item{12.} M. Jacob and P.V. Landshoff, {\it Mod. Phys. Lett.} {\bf A1}, 
657 (1986); 
Ll. Ametller and D. Treleani, {\it Int. J. Mod. Phys.} {\bf A3}, 521 
(1988). 
\item {13.}L.S. Brown Phys. Rev. D5, 748 (1972);
          Y. Akiyama, S. Hori Progr. Theor. Phys. 48, 276 (1973)
\item{14.} G. Calucci and D. Treleani, {\it Nucl. Phys.} {\bf B} (Proc. 
Suppl.) 18C, 187 (1990); 
{\it Int. J. Mod. Phys.} {\bf A6}, 4375 (1991). 
\item{15.} G. Calucci and D. Treleani {\it Phys. Rev.} {\bf D49}, 138 
(1994); 
{\bf D50}, 4703 (1994); J. Bartels and M. W\"{u}sthoff, {\it Z. Phys.}  
{\bf C66}, 157 (1995). 
\item{16.} H.D. Politzer, {\it Nucl. Phys.} {\bf B172}, 349 (1980); 
R.K. Ellis, R. Petronzio and W. Furmanski, {\it ibid.} {\bf B207},  
1 (1981); 
N. Paver and D. Treleani, {\it Nuovo Cimento} {\bf A70}, 
215 (1982); Zeit.  
Phys.{\bf C28}, 187 (1985); B. Humpert, {\it Phys. Lett.}  
{\bf 131B}, 461 (1983); 
B. Humpert and R. Odorico, {\it ibid} {\bf 154B}, 211 (1985);  
T. Sjostrand and 
M. Van Zijl, {\it Phys. Rev.} {\bf D36}, 2019 (1987). 
\item{17.} F. Abe et al., {\it Phys. Rev.} {\bf D47}, 4857 (1993). 
\item{18.} C. Albajar et al. {\it Nucl. Phys.} {\bf B309}, 
405 (1988).
\vfill 
\eject 
 
\par    {\bf Figure captions} 
\vskip .15in 
\item{Fig.1-} The three different kinematical regions which characterize 
semihard hadronic interactions. I: only the single partonic collision, 
described by a single BFKL Pomeron exchange, 
is relevant; II: multiparton collisions are to be taken into 
account, each partonic 
interaction is however well described by a single BFKL Pomeron exchange; 
III: the single BFKL Pomeron exchange is not any more an adequate 
description of the single parton interaction. 
\vskip .15in 
\item{Fig.2-} Cross section for production of minijets  
with rapidity gap as a function of the gap $\Delta y$. The process
is $p\bar{p}$ at $\sqrt{s}=1.8TeV$ and $q_t^{min}=5GeV$. The dotted curve is the 
single scattering term. The continuous and the dashed curves include all 
disconnected multiple parton collisions. Input to the continuous
curve is $\sigma_{eff}=20mb$ and to the dashed curve is
$\sigma_{eff}=12mb$ (see text).
\vfill 
\eject\end  
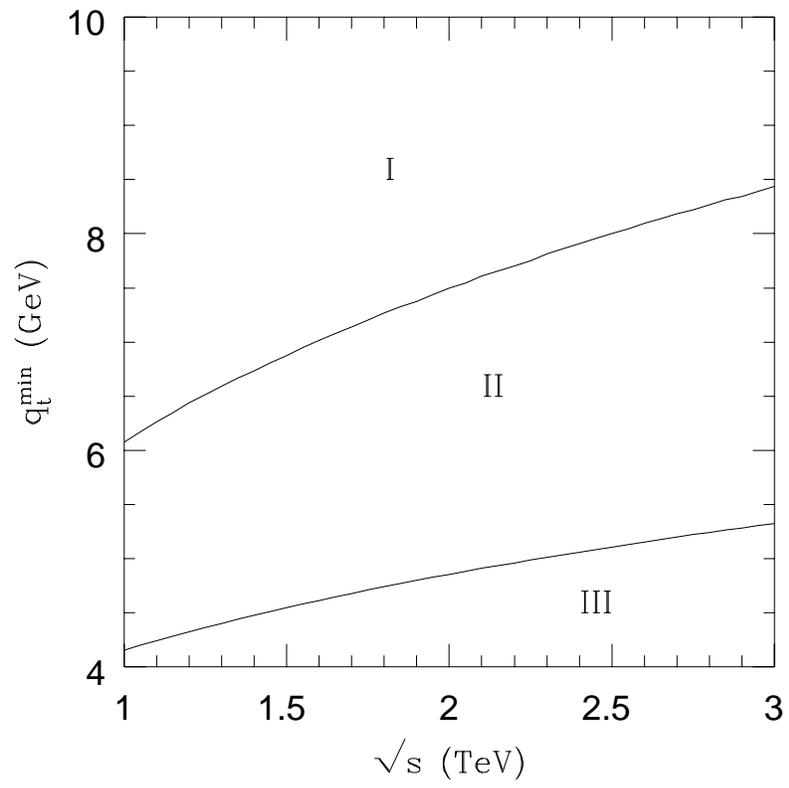
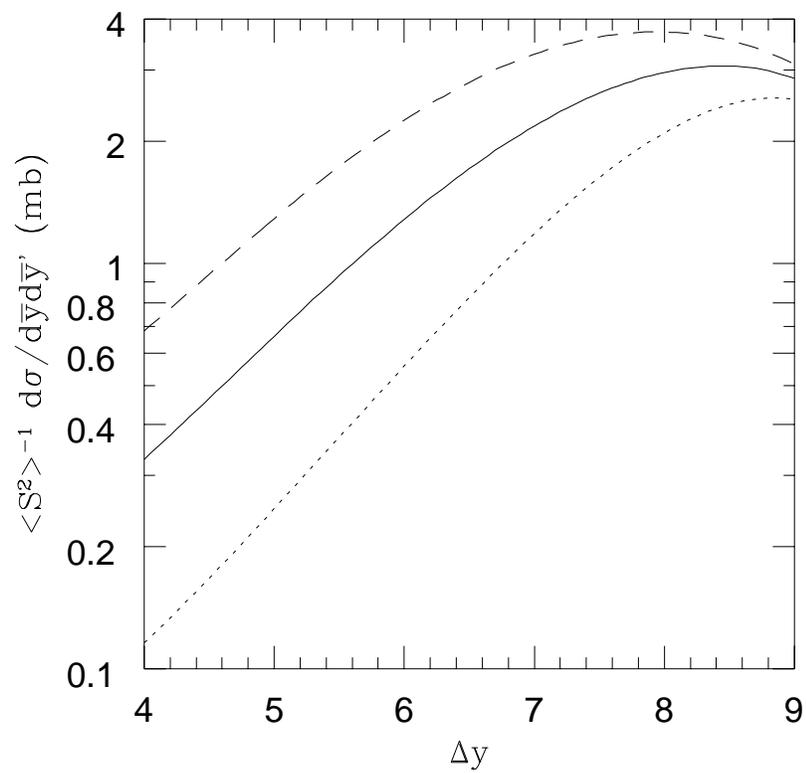
\bye